\RequirePackage{fix-cm}
\documentclass{svjour3}                     

\smartqed  
\usepackage{graphicx}
\usepackage{setspace}
\usepackage{float} 
\usepackage{hyperref}  
\hypersetup{
  colorlinks   = true, 
  urlcolor     = {blue!50!black}, 
  linkcolor    = {blue!50!black}, 
  citecolor   = {red!50!black} 
}
\usepackage[T1]{fontenc} 
\usepackage[utf8]{inputenc} 
\usepackage{amssymb} 
\usepackage{amsfonts}
\usepackage{amsbsy}
\usepackage{soul} 
\setul{}{1pt}  
\setulcolor{red} 
\usepackage{graphicx} 
\graphicspath{{./Figures/}}  
\usepackage{enumitem} 
\usepackage{multirow} 
\usepackage[table]{xcolor} 
\usepackage{colortbl} 
\usepackage{subfig} 
\usepackage{url} 
\usepackage{lineno}
\begin{document}

\title{Experimental Investigation of the Inelastic Response of Pig and Rat Skin under Uniaxial Cyclic Mechanical Loading
}
\titlerunning{Inelastic Response of Skin}        

\author{N. Afsar-Kazerooni \and A. R.  Srinivasa \and J.C. Criscione
}

\authorrunning{Afsar Kazerooni \and Srinivasa \and Criscione} 

\institute{N. Afsar-Kazerooni \at
              Department of Mechanical Engineering, Texas A$\&$M University\\
              College Station, TX, 77843\\
              Tel.: +1979-25-50414\\
              Fax: +1979-84-53081\\
              \email{nazanin26@tamu.edu}           
           \and
          A. Srinivasa \at
             Department of Mechanical Engineering, Texas A$\&$M University\\
             College Station, TX, 77843\\
              Tel.: +1979-86-23999\\
              Fax: +1979-84-53081\\
             \email{asrinivasa@tamu.edu}
              \and
              J. Criscione \at
              Department of Biomedical Engineering, Texas A$\&$M University\\
              College Station, TX, 77843\\
              Tel.: +1979-845-5428\\
              Fax: +1979-845-4450\\
             \email{jccriscione@tamu.edu}
}


\maketitle

\begin{abstract}
Skin is a highly non-linear, anisotropic, rate dependent inelastic, and nearly incompressible material which exhibits substantial hysteresis even under very slow (quasistatic) loading conditions. In this paper, a series of uniaxial cyclic loading tests of porcine and rat skin at different strain rates and with samples oriented in different directions (with respect to the spine) were conducted to study the effect of strain rate and samples orientations with respect to spine on mullins effect and skin inelastic response. A noteworthy feature of skin is that, similar to certain filled rubbers, its mechanical response shifts after the first extension and exhibits softening and hysteresis when loaded under cyclic tension and mullins effect is observed. The results of these strain-controlled cyclic loading tests also indicated that the extent of softening is different for different strain rates and orientations. Also a substantial hysteresis persists even at very low strain rates indicating inelastic behavior beyond the rate sensitive viscoelastic response. Through this series of experiments, by investigating the effect of strain rate on pig skin and rat skin, we conclude that the skin response is rate dependent but inelastic and shows  irreversible changes in fiber orientation which are observed in histology results. Also, skin shows persistent deformation that is only partially recovered even after a long period of unloading. 
\keywords{Mullins Effect \and Skin \and Uniaxial Cyclic Tension testing \and Partially Unloading \and Fiber Orientation and Strain Rate}
\end{abstract}

\section{Introduction}

The largest organ in the body is skin. It is composed of several layers including a stratified, cellular epidermis and an underlying dermis of connective tissue \cite{McGrath10}. More recently a whole new layer of fluid vesicles (called the intersitium) has been discovered by Benias et al. and it exists right below skin surface. There is a layer of fat under the dermis which separates these layers of skin from the rest of the body \cite{Benias18}. The dermis contains collagen, elastin and extra fibrillar matrix materials. The collagen fibers are made of protein and provide the strength of skin and are necessary for the shape and structure of skin \cite{ricard11}. Elastin, which is another component of skin, has a role in mechanical properties of skin too. Oxlund et al. showed that after loading and deformation, elastin fibers play a role in the recoiling mechanism, but they are not primarily responsible for bearing load unlike collagen \cite{OXLUND88}. When these fibers are damaged or lost due to aging, skin loses its stiffness and elasticity and wrinkling is observed \cite{Naylor11}.

Preliminary investigations have shown that these collagenous tissue found in skin, cartilage, blood vessels, and the cornea are tough, compliant, and capable of withstanding multiaxial cyclic loading with minimal damage. Experiments have shown that under loading \cite{LIAO99} and \cite{HILL12}, the fibers are initially slack, then they are gradually “recruited” as deformation proceeds. Eventually the fibers become taut and the stress dramatically increases with very little further elongation. 

Brown et al. showed that stress-strain curve of abdominal human skin under uniaxial tension test contains 3 stages on the curve \cite{Brown73}. Based on the results from scanning electron microscopy, Brown et al. observed that at the beginning of first stage, collagen fibers are wavy and not elongated \cite{Brown73}. At this point, elastin fibers play an important role in stretching and after a while, collagen fibers slowly elongate with increasing strain. This is now responsible for the extension of skin and at this stage the slope of stress strain curve is very low in comparison with the next stages. By further increasing the strain, collagen fibers get aligned. This stretching of collagen fibers at the second stage is known as the linear region \cite{Xu11}. In this stage, collagen fibers are much more aligned in the direction of loading and these aligned fibers increase the stiffness of skin. The stress as well as the stiffness at this stage is high in comparison with the previous stage. At the last stage, the stress reaches an ultimate tensile strength of collagen, the fibers start failing, and the slope decreases again until failure \cite{Brown73} and \cite{Xu11}. 

Haut et al. performed uniaxial tensile testing on rat skin until failure \cite{Haut89}. The results showed that the tensile strength is different when samples are cut in different directions with respect to spine, since they have different direction with respect to Langer lines. Langer lines are correlated to the natural orientation of collagen fibers in the dermis and they are mostly perpendicular to the spine in animals \cite{swaim90}.  Also,  the tensile strength are sensitive to strain rate and show different strength values at high and low strain rates. The tensile strength is highly correlated to the degree of collagen crosslinking \cite{Dombi93}. Annaidh et al. measured the slope of the stress-strain curve for human skin \cite{Annaidh12}. They carried out monotonic uniaxial tensile testing on human skin at different orientations with respect to the Langer lines and histology was done on three samples to find collagen fiber orientation. Slopes for with the angle of 45 degree with respect to Langer lines were between the parallel and perpendicular directions with respect to Langer lines. North et al. used a Schrader double acting hydraulic cylinder (which measures changes in volume by applying pressure), and showed that human abdominal skin can be considered as an incompressible material since changes in volume compressibility of skin is smaller than that of water for large changes in pressure \cite{North78}.

In previous studies, most of the experiments on skin have been only for monotonic loading. While there has been quite a few researches on the inelasticity of soft tissue, there has been very little published on skin under cyclic loading. It has also long been known that when skin undergoes large deformation under cyclic loading, it exhibits strong hysteresis \cite{Fung13}.

Kang et al. investigated the mechanical behavior of porcine skin under uniaxial cyclic loading and from the histological images, they observed irrecoverable micro-mechanisms such as sliding of collagen fibers and fibrils \cite{Kang11}. The experiments and histological studies have clearly demonstrated the inelastic behavior of skin and the mechanisms of damage, however, the modeling of this behavior has been quite challenging. 

Under cyclic loading skin shows some similar response as filled rubbers does. Under these conditions filled rubbers, after loading in the first cycle, most of the softening happens. After the first cycle, the following cycles coincide with each other \cite{Mullins49}. This softening is called mullins effect in elastomers (e.g. \cite{Diani09}). The softening is higher at higher strains. Also, it was observed that at strains lower than the strain at peak stress, and at strains higher than the strain at peak, the softening will disappear and stress-strain response will go back to the first path and follow it. 
Lokshin et al. \cite{lokshin09} have studied the response of precodntioned skin by subjecting it to step strains and fully unloading the sample. They modeleed the skin as purely viscoelastic solid  where the original shape and response  is recovered with time upon unloading. On the other hand,  mullins effect is, in essence, an irreversible shape and response change phenomenon. Purely viscoelastic response functions such as that developed by Lokshin et al. \cite{lokshin09} can not capture this effect and it is necessary to have an inelasticity model that can show this effect \cite{lokshin09}. Furthermore, Nava et al. \cite{Nava04} demonstrated that biological tissues during surgery behave more like tissue during the first cycle under cyclic loading rather than the tissue after preconditioning. They further go on to  state that ''This analysis demonstrates that a quasi-linear viscoelastic model fails in describing the observed evolution from the "virgin" to the preconditioned state. Good agreement between simulation and measurement are obtained by introducing an internal variable changing according to an evolution equation''. Lanir et al. \cite{lokshin09}. Primarily focused on "precondioned" skin and therefore did not account for the inelastic changes. They also did not report on the measured response during preconditioning, focusing instead only on the stabilized response. In view of this, in this paper we present experimental results on the response of ``virgin" skin by subjecting it strain cycling at different rates, with  both full and partial unloading. It is hoped that the results will be of use to modelers of skin for use in  surgical simulators or virtual reality simulators, diagnosis as well as simulation of the interaction of skin with wearable devices \cite{Avis00} and \cite{Satava99}. 

There are only a  few studies on mullins effect on skin. In terms of the experimental evidence Munoz et al.\cite{Munoz08} performed uniaxial cyclic loading on abdominal region of mice skin . The results show softening after first cycle and they expressed it as mullins effect, similar to the case of the response of filled rubber. Several features of the response are similar to the response in this work which will be discussed in next sections. However, they have restricted their loading to only one direction with respect to the spine. Furthermore, since it is not clear from their paper what kind of stress and strain measures were used, it is difficult to have a detailed comparisons with models etc. More Recently, Zhu et al. \cite{Zhu14} performed monotonic and force controlled cyclic tension tests on the dorsal area of pig skin from two perpendicular direction . Since, they ran force controlled tests, it is difficult to see internal loops and measure the extent of softening.
There are still many unanswered questions regarding the inelastic response of skin. For example, is the response elastic under quasistatic loading so that hysteresis is entirely due to viscoelasticty? Is there a directional dependence of the response? If so, are they qualitatively similar? What happens when skin is reloaded without being fully unloaded, does it follow the unloading curve or does it follow an entirely new path?
The aim of this paper is to show that skin shows a strong mullins effect, i.e., irreversible change in response superimposed on a time dependent phenomenon and to quantify the effect in terms of both stress strain response as well as fiber reorientation (histological data) so that modelers can utilize the data. The fact that there is an irreversible change is demonstrated both via the stress strain response and histology.

For this purpose, the experimental data is provided on the response of pig and rat skin subjected to strain controlled partial cyclic loading at different strain rates and with three different directions compared to the spine axis followed by images from optical microscopy to show fiber alignment before and after cyclic tensile testing. The experiments demonstrate that both pig and rat skin have complex behavior (especially under partial unloading) but show strikingly similar features in common with the response of filled rubbers although the microstructural processes that give rise to this are different. The experiments also reveal that there is strong anisotropy of response (even though the response shows a similar nature in different directions). 
We show further that the internal hysteresis decreases with increasing strain rates and that hysteresis persists for quasistatic response (which is a mark of inelasticity)  . We quantify the change in the hysteresis by measuring the amount of energy dissipated per unit strain for different strain rates.
\section{Material and Method}
\subsection{Sample Preparation and Set-up}
\begin{figure}
\centering
\includegraphics[scale=1, trim =0 0 0 0,clip]{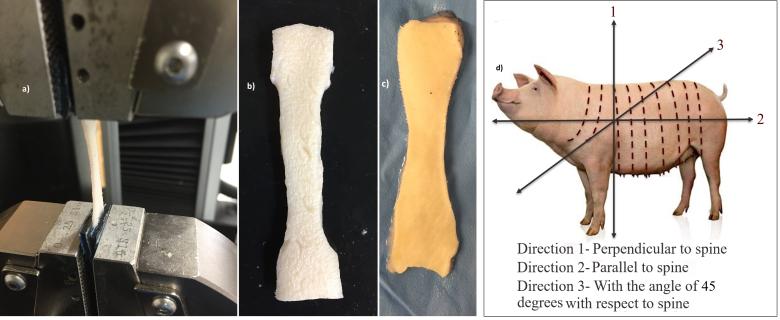}
\caption{a) Pig skin sample in wedge action grips, b) The pig skin dog bone specimen, c) The rat skin dog bone specimen and d) Three different directions that samples were cut, direction 1 is perpendicular to spine, direction 2 is parallel to spine and direction 3 is with the angle of 45 degrees with respect to spine.}
\label{fig:1}
\end{figure}

Large skins from the abdominal and back region of adult pigs were obtained from Veterinary School of Texas A$\&$M University. 
First, skins were shaved clean of hair using a razor blade and 12 samples with length of 20 mm, average thickness of 2 mm and average width of 7 mm were cut from the large skin with epidermis and dermis layers (the initial widths and the initial thicknesses were measured at four different location along the length and then their average were reported as an initial width and thickness). Samples were cut in three different directions with respect to the direction of spine, perpendicular, parallel and with the angle of 45 degrees with respect to spine to investigate the effect of anisotropy of skin (\autoref{fig:1}-(d)) on behavior of skin under uniaxial cyclic loading. In order to prevent slipping of samples from grips, sand paper was glued to a piece of rubber and placed between the skin sample and the wedge action grips of Instron 5567 testing machine with load cell capacity of $\pm$ 5 kN (\autoref{fig:1}-(a)). Also, in order to study the behavior of different kind of skin, 10 rat skin samples were harvested from  adult rats from Veterinary School of Texas A$\&$M University with length of 20 mm, average thickness of 1.5 mm and average width of 5 mm. Samples preparation and gripping for rat skins were the same as pig skin. 
In order to find the relation between actual tissue deformation and grip displacement, after calculating and comparing the strain from grip displacement, and the strain from image correlation along the sample from different positions of the sample, the difference between both measurement was observed to be less than 5 percent. The details of the  comparison between grip displacements and DIC is described in the Appendix.
\subsection{Experiments} 
\begin{figure}
\centering
\includegraphics[scale=0.8, trim =0 0 0 0,clip]{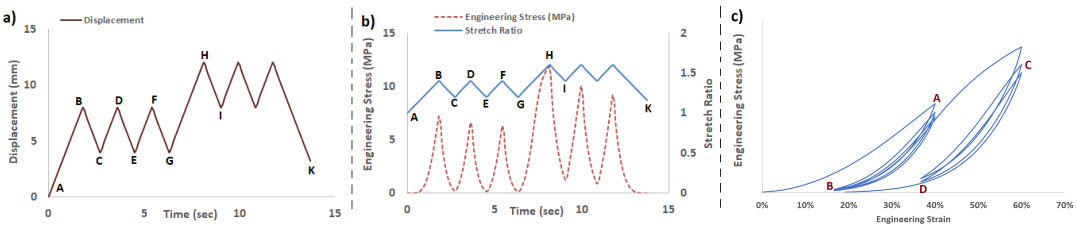}
\caption{a) Displacement path of uniaxial partially loading and unloading tension testing for pig skin at the strain rate of 0.22 $s^{-1}$ and similar path was used for rat skin, b) The stress and strain path of uniaxial partially loading and unloading tension testing of pig skin which was cut perpendicular to the direction of spine (direction 1) at the strain rate of 0.22 $s^{-1}$. Note that positions of the peak and valley coincide showing that, at this slow rate, there is no lag. Similar graphs of the raw data available for the other directions and for the rat skin which has the same features and response as this pig skin, c) Schematic engineering stress-engineering curve for pig skin, A-B shows the three cycles until the maximum strain of 40$\%$ and C-D shows the next two cycles until the maximum strain of 60$\%$.}
\label{fig:2}
\end{figure}

Pig skin samples were tested under cyclic uniaxial loading at strain rates of 0.011 $s^{-1}$ (displacement rate of 13 mm/min) and 0.22 $s^{-1}$ (displacement rate of 260 mm/min) at the room temperature. Similar tests were done on rat skin samples at the strain rate of 0.008 $s^{-1}$ (displacement rate of 10 mm/min) and 0.5 $s^{-1}$ (displacement rate of 500 mm/min) also at the room temperature. 
In order to keep the samples wet during the tension tests and prevent them from drying that it does not affect or change the behavior of skin \cite{Berardesca95}, Veterinary 0.9 percent Sodium Chloride solution was sprayed on each sample during the test. The cyclic loading protocol was as follows:  first, samples were stretched to a specific elongation (point B in \autoref{fig:2}-(a)) and then partially unloaded until reached a certain elongation (point C in \autoref{fig:2}-(a)). Then this strain controlled loading and unloading cycles were repeated for a few times and force was measured. 
For each direction, two samples were tested to check the repeatability of the tests. In order to investigate the effect of time on recovery of skin, one sample from rat skin and one sample from pig skin were tested at day 0 and again after 3 days of keeping at refrigerator at $0^{\circ}C$ the same samples were tested. Also, the morphology of outer surface of skin (which is called stratum corneum of the same sample) was observed under the optical microscope to study changes that happened after 3 days.
Finally, in order to find fiber orientations, after cyclic loading, pig skin samples were cut and after fixation, tissue processing and sectioning, they were stained. Then slides were observed under VHX-600 Digital Microscope. Fiber orientations were measured by ImageJ software.

\section{Results}

Engineering stress and engineering strain were computed from force displacement data. Engineering stress was calculated by dividing the force over the reference area (initial thickness multiply by initial length). Engineering strain was computed by dividing displacement over initial length. \autoref{fig:2}-(b) shows the path of engineering stress and engineering strain versus time during partially loading and unloading.

\begin{figure} [ht]
\centering
\includegraphics[scale=1, trim =0 0 0 0,clip]{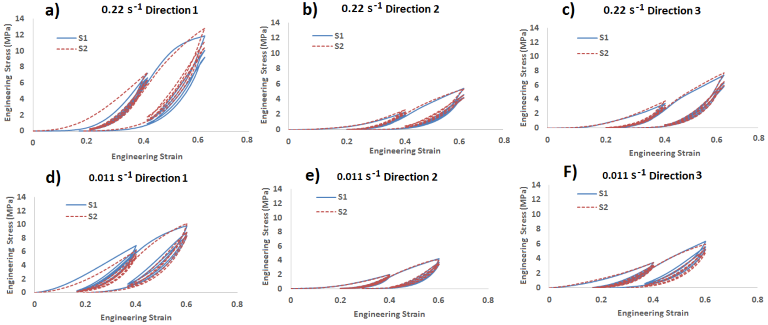}
\caption{Cyclic uniaxial tension testing of pig skin  at different directions. Also, it shows  repeatability in each direction a) Direction 1 which is perpendicular to the spine direction at the strain rate of 0.22 $s^{-1}$, b) Direction 2 which is parallel to the spine direction at strain rates of 0.22 $s^{-1}$, c) Direction 3 which is cut by the angle of 45 degree with respect to spine at the strain rate of 0.22 $s^{-1}$, d) Direction 1 which is perpendicular to the spine direction at the strain rate of 0.011 $s^{-1}$, e) Direction 2 which is parallel to the spine direction at the strain rate of 0.011 $s^{-1}$ and f) Direction 3 which is cut by the angle of 45 degree with respect to spine at the strain rate of 0.011 $s^{-1}$.}
\label{fig:3}
\end{figure}

\autoref{fig:3} shows the stress strain response of pig skins. The cyclic uniaxial testing of pig skin dogbone samples which were cut in directions of perpendicular, parallel, and with an angel of 45 degrees with respect to the spine at the strain rate of 0.22 $s^{-1}$ are shown in \autoref{fig:3} (a)-(c). Tests were found to be repeatable and followed the same path and showed a repeatable response within the margin of less than 8 percent. Same experiments were carried out at the strain rate of 0.011 $s^{-1}$ and the results are shown in the \autoref{fig:3} (d)-(f), these figures show repeatability of the test.

\begin{figure} [ht]
\centering
\includegraphics[scale=1, trim =0 0 0 0,clip]{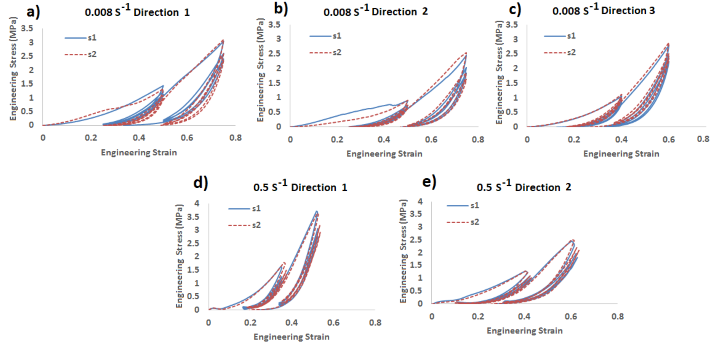}
\caption{Cyclic uniaxial tension testing of rat skin at different directions. Also, it shows repeatability in each direction a) Direction 1 which is perpendicular to the spine direction at the strain rate of 0.008 $s^{-1}$, b) Direction 2 which is parallel to the spine direction at the strain rate of 0.008 $s^{-1}$, c) Direction 3 which is cut by the angle of 45 degree with respect to spine at the strain rate of 0.008 $s^{-1}$, d) Direction 1 which is perpendicular to the spine direction at the strain rate of 0.5 $s^{-1}$ and e) Direction 2 which is parallel to the spine direction at the strain rate of 0.5 $s^{-1}$.}
\label{fig:5}
\end{figure}
\begin{figure}
\centering
\includegraphics[scale= 1, trim =0 0 0 0,clip]{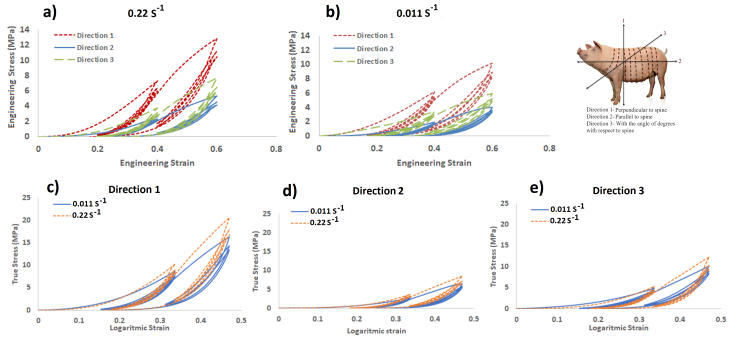}
\caption{The effect of samples direction with respect to spine (a-b): comparison of response of pig skin under uniaxial cyclic loading at direction 1 which is perpendicular to the spine direction, direction 2 which is parallel to the spine direction and direction 3 which is cut by the angle of 45 degree with respect to spine a) the strain rate of 0.22 $s^{-1}$ and b)the strain rate of 0.011 $s^{-1}$. The effect of strain rate on pig skin samples (c-e): comparison of true stress and logarithmic strain of pig skin under uniaxial cyclic loading c) direction 1  which is perpendicular to the spine direction at the strain rates of 0.22 $s^{-1}$ and 0.011 $s^{-1}$, d) direction 2 which is parallel to the spine direction at the strain rates of 0.22 $s^{-1}$ and 0.011 $s^{-1}$, and e) direction 3 which is cut by the angle of 45 degree with respect to spine at the strain rates of 0.22 $s^{-1}$ and 0.011 $s^{-1}$.}
\label{fig:4}
\end{figure}

\autoref{fig:5} shows the results of uniaxial cyclic partially loading and unloading of rat skin at two different strain rates of 0.008 $s^{-1}$ and 0.5 $s^{-1}$. At the strain rate of 0.008 $s^{-1}$ samples were tested from three different orientations with respect to the spine: direction 1 which is perpendicular to the spine direction, direction 2 which is parallel to the spine direction and direction 3 which is cut by an angle of 45 degree with respect to spine. At the strain rate of 0.5 $s^{-1}$ samples were tested from two different orientations, direction 1 which is perpendicular to the spine direction and direction 2 which is parallel to the spine direction.
The stress strain response under cyclic loading for both pig and rat skin shows the following features:
\begin{enumerate}
    \item During the first loading (AB in \autoref{fig:2}-(a)) the initial modulus is very small but increases rapidly beyond a certain strain.
    \item When the strain is reduced (BC in \autoref{fig:2}-(a)) the unloading curve shows evidence of hysteresis.
    \item Upon reloading, the response approximates the unloading curve (but not exactly, i.e. it shows unloading reloading hysteresis) up to nearly the previous peak and then appears to continue along the original source this is usually referred to as softening.
    \item The hysteresis between the first loading and unloading (i.e. path ABC) is much greater than the subsequent cycles (i.e. path CDEFG). 
    \item All of this clearly reveals mullins effect in skin.
    \item If the unloading curve reaches zero load (i.e. full unloading) the response reveals a persistent strain (see point K in \autoref{fig:2}-(a)). 
    \item Comparing \autoref{fig:3}(a)-(c) and \autoref{fig:3}(d)-(f), we see that if the same experiment is carried out at a higher rate, the stresses are higher at a certain strain (i.e. there is a certain amount of strain rate dependent hardening).
    \item However, the unloading, reloading hysteresis decreases with increasing rate (\autoref{tab:3}).
    \item Also, at higher strain rates, the transition from lower modulus to higher modulus and hardening region happens at lower strain a confirming observations by Zhou et al. \cite{Zhou10}.
    \item Next, upon comparing the response between the 0 and 90 direction, the direction perpendicular with respect to spine has higher slope (especially at region 2 and 3) than the parallel direction (\autoref{fig:4}(a)-(b)).
    \item The 45 degree experiment shows that it has the slope between the other two directions (\autoref{fig:4}(a)-(b)).
    \item  True stresses and logarithmic strains at each direction at different strain rate (\autoref{fig:4}(c)-(e) shows similar behavior as engineering stresses and strains. 
    \item The behavior of the rat skin (see \autoref{fig:5}) also showed similar qualitative features although the stress amplitudes were quite different.
\end{enumerate}

\begin{figure}
\centering
\includegraphics[scale=1, trim =0 0 0 0,clip]{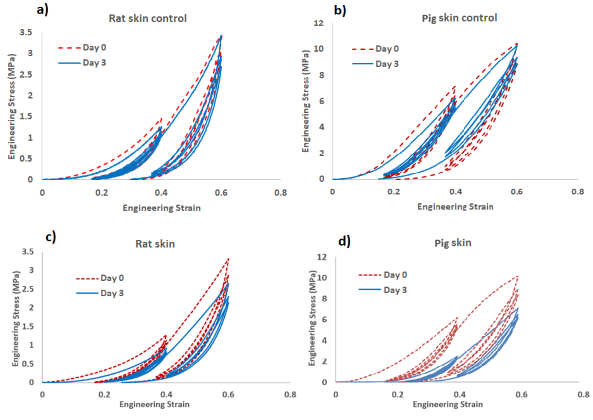}
\caption{Uniaxial cyclic loading testing until 60 percent strain at the strain rate of 0.011 $s^{-1}$. Samples were cut in direction 1 which is perpendicular to the spine direction at day 0 and day 3 a) Previously untreated rat skin samples at day 0 and day 3 as control samples, b) Previously untreated pig skin samples at day 0 and day 3 as control samples, c) A rat skin sample at day 0 and retested at day 3 and d) A pig skin sample at day 0 and retested at day 3, at both days skin shows similar response. However, the retested samples show substantial softening due to damage after cyclic testing.}
\label{fig:6}
\end{figure}

Skin shows anisotropic behavior \cite{Minns73}. This behavior of skin was studied under uniaxial cyclic loading on pig skin at different strain rates and different samples from 3 different directions with respect to spine. \autoref{fig:4}(a)-(b) show anisotropy in pig skin behavior. \autoref{fig:4}(c)-(e) shows that at the effect of strain rate is more dominant at higher strains.  
In order to investigate whether these are permanent or not, one sample of pig skin and one of rat skin were subject to uniaixal cycling upto a maximum strain of 60 $\%$ (see \autoref{fig:6} at a strain rate of 0.011 $s^{-1}$ on both rat skin and pig skin with the original orientation of perpendicular to the orientation of spine. Both samples were tested at day 0 and then kept in a refrigerator (at $0^{\circ}C$) for 3 days. After 3 days, uniaxial testing were carried out on them again. In order to investigate the effect of storing skin on mechanical behavior of skin, one previously untreated rat and one pig samples were tested after 3 days, as control samples.

\begin{figure} 
\centering
\includegraphics[scale=1, trim =0 0 0 0,clip]{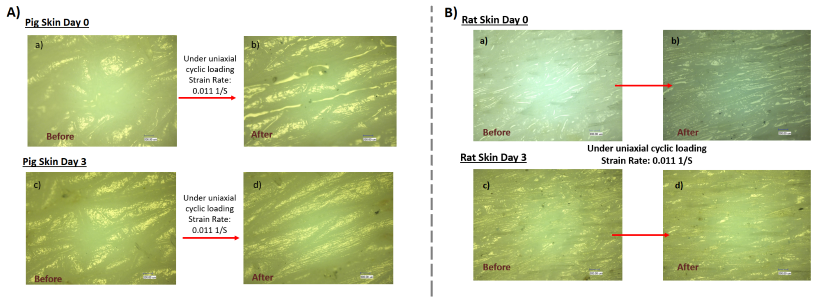}
\caption{A) Optical microscopic images of the surface of outer layer of pig skin dog-bones under cyclic uniaxial tension testing at the strain rate of 0.011 $s^{-1}$ shows fiber directions A-a) Previously untreated sample, before cyclic tension test at day 0, A-b) Previously untreated sample, after cyclic tension test until 60 percent strain at day 0, A-c) Previously treated sample, before cyclic tension test at day 3 and A-d) Previously treated sample, after cyclic tension test until 60 percent strain at day 3, B) Optical microscopic images of the surface of outer layer of rat skin dog-bones under cyclic uniaxial tension testing at the strain rate of 0.011 $s^{-1}$ shows fiber directions B-a) Previously untreated sample, before cyclic tension test at day 0, B-b) Previously untreated sample, after cyclic tension test until 60 percent strain at day 0, B-c) Previously treated sample, before cyclic tension test at day 3 and B-d) Previously treated sample, after cyclic tension test until 60 percent strain at day 3.}
\label{fig:7}
\end{figure}

\autoref{fig:7} shows images of the surface of outer layer of pig skin and rat skin which is called stratum corneum under optical microscopy. This layer is composed of dead and dying skin cells from the underlying epidermis. \autoref{fig:7}(A-a) shows the pig skin image before uniaxal cyclic testing at day 0 and \autoref{fig:7}(A-b) shows the pig skin image after uniaxial cyclic loading. \autoref{fig:7}(A-c) shows the image of the pig skin after cyclic loading and storing in a fridge for 3 days. \autoref{fig:7}(A-d) shows pig skin image after cyclic loading at day 3 and it is elongated. \autoref{fig:7}(B-a) is rat skin before uniaxial cyclic loading at day 0, as the figure shows, fibers have random orientation, and in \autoref{fig:7}(B-b), fibers are elongated. Then after storing in a fridge for 3 days, the sample was investigated under microscope. As \autoref{fig:7}(B-c) shows, fibers are still elongated in comparison with \autoref{fig:7}(B-a), and \autoref{fig:7}(B-d) shows the rat skin image after uniaxial cyclic testing after 3 days.

\begin{figure}  [ht]
    \centering
    \includegraphics[scale=0.9]{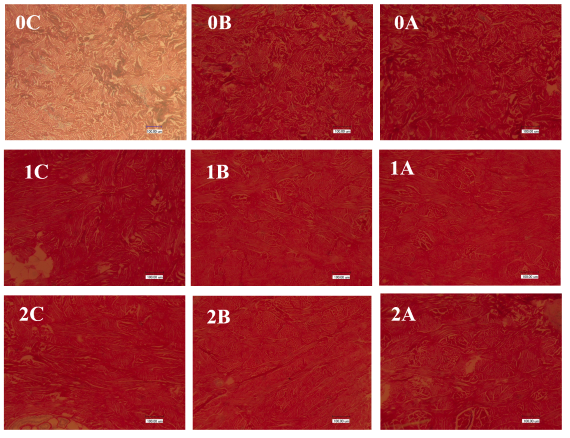}
    \caption{ Microstructure of samples after staining: before deformation (sample 0) and after deformation under uniaxial tensile testing: sample 1 was cut in the perpendicular direction with respect to spine and 2 was cut in the parallel direction with respect to spine.}
    \label{fig:8}
    \end{figure}
    
     \begin{figure} 
    \centering
    \includegraphics[scale=1]{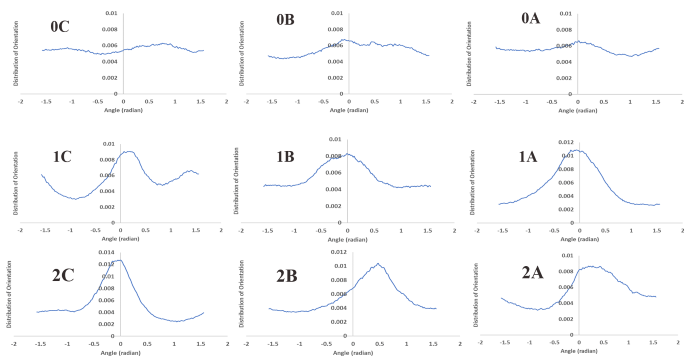}
    \caption{Fiber distributions of control sample with no deformation (sample 0) and fiber distributions after uniaxial cyclic testing: sample 1 was cut in the perpendicular direction with respect to spine and 2 was cut in the parallel direction with respect to spine}.
    \label{fig:9}
    \end{figure}
    
    \begin{figure} 
    \centering
    \includegraphics[scale=0.9]{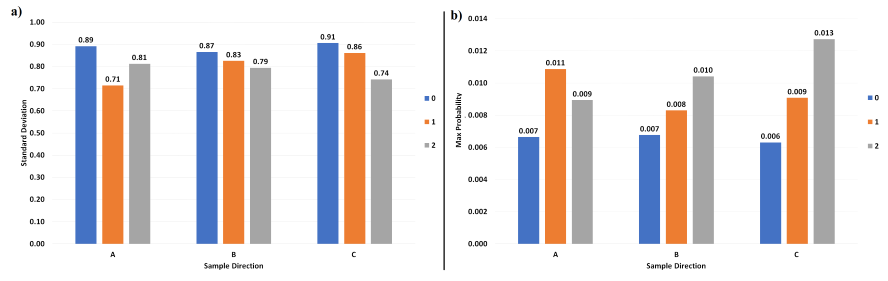}
    \caption{a) Standard deviation over fiber distribution and b) Maximum probability of fiber distributions of each sample at different directions.}
    \label{fig:10}
    \end{figure} 
Fiber recruitments and realignment of collagen fibers plays an important role in mechanical behavior of biological tissues. Histology has been done to help modelers to have a more comprehensive data to model the behavior of skin. \autoref{fig:8} shows the histology results of pig skin samples before and after deformation. Sample 0 is an undeformed sample and it was cut in 3 different direction: A is a cut in a perpendicular direction with respect to spine, B is a cut in a parallel direction with respect to spine and C is a cut parallel to dermis. A and B are cuts through all layers. Sample 1 was cut in the perpendicular direction with respect to spine and 2 was cut in the parallel direction with respect to spine. For these two samples, A is a cut in a parallel direction with loading direction, B is a cut in the perpendicular direction with loading direction and C is a cut parallel to dermis. \autoref{fig:10}-a shows Standard deviation over fiber distribution and \autoref{fig:10}-b shows Maximum probability of fiber distributions of each sample at different directions. Quantitative experimental data was provided on the distribution of fibers before and after deformation. These fibers recruitment and reorientation were quantitatively measured in \autoref{fig:9}.
     
\section {Discussion}
Usually experiments in the literature are about fully unloading conditions \cite{lokshin09}. In this paper, partially unloading was performed (\autoref{fig:3} and \autoref{fig:5}) which is not usually done, and it reveals information about structure of materials. In fully loading and unloading, the hysteresis is larger than partially unloading, so it is necessary to perform both types of unloading to find a more comprehensive constitutive equation that can capture the whole behavior of skin. One of the focuses of this study is to show the effect of strain rate on mullins effect (the softening that happens during the first unloading). As Eshel et al. showed, the relaxation time for the tissue is in the order of hundreds of seconds \cite{Eshel01}. When strain rate is fast, it means that strain rates are faster than relaxation time for the tissue. For example, a strain rate of 0.1 per second \cite {Eshel01} (to a strain of 0.6 implies that the experiment will be completed in 6 seconds is very fast as compared to the relaxation time of the tissue. On the other hand, 0.008 per second will be considered slow since the time taken to complete the loading to 0.6 will be about the relaxation time, ie the material will have time to relax before the experiment is complete so it is considered slow. 
Skin shows anisotropic behavior \cite{Minns73}. This behavior of skin was studied under uniaxial cyclic loading on pig skin at different strain rates and different samples from 3 different directions with respect to spine. \autoref{fig:4} compares skin responses at different direction and then compares different strain rates at each direction and it shows that the effect of strain rate is more dominant at higher strains. Orientation of samples with respect to Langer lines or spine affects the mechanical properties of skin and stress strain curves; especially modulus at the second stage will be different for each orientation. Stiffening is increased by increasing the angle between the sample and the orientation of spine. The maximum value is observed for the direction of perpendicular to the spine which is almost parallel to Langer lines of that region \cite{Ottenio15}. 
As it was expected, \autoref{fig:5} shows that the results of rat skin were similar to the results of the pig skins. In their stress strain curves, mullins effect were observed, they showed that softening occurs after the first cycle and the amount of softening is increased by increasing the strain \cite{Mullin48}. Also, they showed anisotropic behavior similar to pig skin and in direction perpendicular to the spine, rat skin is more stiff and by deceasing the angle between the direction of sample and the spine, stiffness decreases \cite{Ottenio15} but pig skin is more sensitive to direction with respect to spine than rat skin.
In order to investigate the effect of strain rate and direction on the mullins effect and softening that happens to skin, maximum stresses at both first and second cycles are normalized to the maximum stress at first cycle. \autoref{tab:1} shows that stress at the second cycle approximately decreases by 11 percent for the strain rate of 0.011 $s^{-1}$. For the strain rate of 0.22 $s^{-1}$, softening happened by 15 percent. The results show that the amount of softening is more sensitive to strain rate than orientation of samples with respect to spine or Langer lines. The stress strain behavior dependency on strain rate is due to the movement of collagen fibers which are starting to be aligned and uncoiled \cite{lanir79}. \autoref{tab:2} expresses similar behaviors between pig skin and rat skin. It shows that rat skin is more sensitive to strain rate rather than orientation with respect to the spine orientation. At low strain rate, the amount of softening was approximately 13 percent and at higher strain rate (0.5 $s^{-1}$), it was approximately 20 percent. 
\begin{table}
\caption{Normalized stresses at the end of each cycle at the strain of 0.6 at different strain rates on pig skins which were cut in direction 1 which is perpendicular to the spine direction, direction 2 which is parallel to the spine direction and direction 3 which is cut by the angle of 45 degree with respect to spine. This table shows the amount of reduction in stress after the first cycle.}
\small
\label{tab:1}
\centering
\begin{tabular}{c|c|c|c|c|c|c}
\hline
\multirow{3}{0.1em}{} & Direc. 1 & Direc. 1 & Direc. 2 & Direc. 2 & Direc. 3 & Direc. 3\\
Normalized  & 0.011  & 0.22 & 0.011 & 0.22  & 0.011 & 0.22 \\
stress & ($s^{-1}$) &($s^{-1}$) & ($s^{-1}$) &($s^{-1}$) & ($s^{-1}$)& ($s^{-1}$)\\
\hline
First cycle & 1 & 1 & 1 & 1 & 1 & 1 \\
\hline
Second cycle & 0.89 & 0.86 & 0.89 & 0.85 &  0.88 & 0.85 \\
\hline
\end{tabular}
\end{table}

\begin{table} 
\caption{Normalized stresses at the end of each cycle at the strain of 0.5 and 0.008 $s^{-1}$ at different strain rates on rat skins which were cut in direction 1 which is perpendicular to the spine direction, direction 2 which is parallel to the spine direction.}
\small
\label{tab:2}
\centering
\begin{tabular}{c|c|c|c|c}
\hline
\multirow{3}{0.1em}{} & Direc. 1 & Direc. 1 & Direc. 2 & Direc. 2 \\
Normalized & 0.008 & 0.5 ($s^{-1}$)& 0.008 ($s^{-1}$) & 0.5 ($s^{-1}$) \\
stress &  ($s^{-1}$) & ($s^{-1}$) & ($s^{-1}$) & ($s^{-1}$)\\
\hline
First cycle & 1 & 1 & 1 & 1 \\
\hline
Second cycle & 0.86 & 0.80 & 0.87 & 0.82 \\
\hline
\end{tabular}
\end{table}

\subsection{Persistent strain and original response (almost) recovers after unloading for a few days} 
As noted in the general observations, the stress strain cyclic loading graphs reveal that the samples do not fully recover to their original shapes upon unloading after cyclic loading. In this paper we do not investigate the effect of temperature on behavior of skin, however Xu et al. investigated the effect of temperature on behavior of skin and they showed that by increasing the temperature, skin is damaged \cite{XU08}. However, \autoref{fig:6} shows that after few days while the response was SIMILAR in shape, the actual stresses were much lower, i.e. there was substantial softening. One of the reason for this softening may be the permanent damage that may happen to collagen fibers during the cyclic loading and permanent stretches (3 mm) were observed on both rat and pig skin samples which did not recover after 3 days.
Lokshin et al. in their work always did abrupt straining up to a particular level of strain, holding fixed and fully abrupt unloading\cite{lokshin09}. This kind of loading program mostly highlights viscoelasticity. They have actually modeled skin as a purely viscoelastic material and they have not considered weather or not there is a permanent deformation. The aim of our paper is to demonstrate with a combination of stress strain response as well as histology that there is an irreversible damage in the tissue. In their work\cite{lokshin09}, the strain was up to 30 percent. At lower strains (such as 30 percent), mullins effect and permanent deformation are not easily identifiable. However, at a strain around 60 percent like what we did, mullins effect can be observed easily. \autoref{fig:6} shows that even after 3 days, the material did not recover. On the other hand, the viscoelastic model\cite{lokshin09} is recoverable and will be back to its original response. 
Residual strains were observed after unloading due to changes in tissues structure \cite{sellaro07}. Actually, the inelastic behavior during mechanical testing is due to the change in microstructure. Fiber recruitments and realignment of collagen fibers plays an important role in mechanical behavior of biological tissues. After uniaxil tensile testing, some of the fibers do not go back to their original configuration and they remain in a straight state which is their deformed position which leads to residual strains in the tissue after unloading. Also, by comparing the results between rat skin and pig skin, the amount of softening observed on pig skin after 3 days is approximately 30 percent (obtained by comparing the maximum stresses at the maximum strain at day 0 and day 3), which is more than the softening of rat skin which was approximately 20 percent. It must be added that control samples (previously untreated samples) show the same stress-strain curve with a negligible decrease in stress value, so it can be said that the softening is due to the cyclic loading, not storing for 3 days. 
By comparing pictures in \autoref{fig:7}, it is observed that after unixial loading, cells are elongated. Also, after cyclic loading and storing in a fridge for 3 days, fibers are still elongated with respect to undeformed state. These observations helped us to model the behavior of skin \cite{Afsar18}. Deformation mechanisms start with randomly orientated fibers with low stiffening, then fiber recruitment starts and stiffening increases, finally fiber slippage starts and competes with fiber recruitment until fiber tearing causes permanent damages. If unloaded before tearing, partial recovery is achieved \cite{Afsar18}.
 Histology results (\autoref{fig:8}) show that microstructure of skin undergoes permanent changes. Compared to control samples which fibers are randomly oriented, after uniaxial testing, fibers were mostly oriented in one direction. 
 As \autoref{fig:10}-a shows for the control sample in each cut orientation, standard deviation is the highest, which interprets the orientation distribution is wide and there is no preferred direction and it shows randomly distribution. However, the results show that after uniaxial tensile testing there is a preferred orientation in fiber distributions and it shows the alteration mechanisms, fiber recruitment and the inelastic fiber straightening.
\subsection{Quantification of hysteresis at different strain rate}
The key difference between inelasticity and viscoelasticity has to do with the persistence of hystersis at quasi-static strain rates. In viscoelasticity, at very low strain rates, there is no hysteresis. However, here at very low strain rates, hysteresis still exists, which is a marker for inelasticity \cite{srinivasa09} and \cite{rajagopal16}. Also, after 3 days there is a permanent deformation which did not recover. In viscoelastic material, recovery occurs. To investigate the inelastic response of skin, it is necessary to measure the area of hystereses to find the amount of dissipated energy. Since the total work is $\int\sigma d\varepsilon$, it is possible to approximate this by Trapezoidal rule from the experimental data. The area of internal loops for pig skin samples (at strain rates of  0.22 $s^{-1}$ and 0.011 $s^{-1}$ ) an rat skin sample (at strain rate of  0.011 $s^{-1}$) which were cut in direction 1 (perpendicular to spine) were measured from A to B in \autoref{fig:2}-(c) at the maximum engineering strain of 40$\%$. The next two cycles were calculated from C to D in \autoref{fig:2}-(c) at the maximum engineering strain of 60$\%$. The calculated areas of internal loops which show the amount of dissipated energy are reported in \autoref{tab:3}. The amount of dissipated energy at low strains (A-B) for both strain rates are very close and they are lower than the amount of dissipated energy at high strains (C-D). Also, by increasing the strain (C-D), the amount of dissipated energy at lower strain rate (0.011 $s^{-1}$) is more  than that at higher strain rate (0.22 $s^{-1}$) which indicates that by increasing the strain rate, there is a decrease in hystersis. At each strain rate, the hysteresis gets larger by increasing the maximum strain. For skin samples, the amount of dissipated energy at the third region of stress-strain curve is more than at the second region. The amount of dissipated energy for rat skin is less than  the amount of dissipated energy for pig skin at the same strain rate. Also, in rat skin, the  dissipated energy is less sensitive to the maximum strain compared to pig skin. For the rate skin, at very high strain rate (0.5/sec), since the next loading curve is lower than the previous unloading curve, the area between these two curves is negative although the magnitude is higher, indicating a stronger softening behavior during reloading  at higher rates. 
\begin{table} 
\caption{Comparison of the area of internal loops of pig skin samples and rat skin samples which were cut in direction 1 which is perpendicular to the spine direction at different strain rates at the maximum engineering strain of 40$\%$ and 60$\%$.}
\small
\label{tab:3}
\centering
\begin{tabular}{c|c|c|c|c|c}
\multirow{3}{0.1em}{} Type of & Maximum & Strain & \multicolumn{3}{c}{Dissipated Energy ($\frac{MJ}{m^3}$)}\\
Skin & Eng. Strain & Rate  $(s^{-1})$  & Cycle 1 & Cycle 2 & Cycle 3\\
\hline
Rat & 40 $\%$ & 0.011 & 0.013  & 0.014 & 0.014 \\

Pig & 40 $\%$ & 0.011 & 0.05 & 0.05 & 0.06\\

Rat & 60 $\%$ & 0.011 & 0.035 & 0.041 & N/A \\

 Pig & 60 $\%$ & 0.011 & 0.3 & 0.32 & N/A\\
\hline
Pig & 40 $\%$ & 0.22 & 0.08 & 0.09 & 0.12\\

Pig & 60 $\%$ & 0.22 & 0.18 & 0.21 & N/A \\
\hline 
Rat & 40 $\%$ & 0.5 & -0.05 & -0.05 & -0.01 \\

Rat & 60 $\%$ & 0.5 & -1.39 & -1.31 & N/A \\
\hline

\end{tabular}
\end{table}
\section{Conclusions}
We showed that under uniaxial cyclic loading test, skin shows hysteretic response, and mullins effect was observed. Upon investigating the effect of strain rate and comparing results of tests performed at different strain rates, softening was observed. The softening shows sensitivity to strain rate. At higher strain rates, mullins effect is more significant. Orientation of samples with respect to Langer lines or spine affects the mechanical properties of skin and stress strain curves: stiffening increases by increasing the angle between the sample and the orientation of spine. However, the extent of softening is more related to strain rate rather than sample orientation with respect to spine. Since at very low strain rates hystereses still exist, the skin behavior is more inelastic than viscoelastic. In addition to stress-strain curves, optical microscopic images helped us to a better understanding of skin behavior under cyclic loading. These images show fibers realignment which causes stiffening at higher strains. There is also fiber slippage at high stresses. These two phenomena are responsible for behavior of skin until rupture which is a permanent damage. Despite of all limitations during this study, like sample sizes or temperature effect on skin behavior, we characterized the skin behavior under cyclic loading. Also, we have done histologies and quantified density of fiber distributions which help modelers to have a comprehensive data to model the behavior of skin. Our study does not precondition the material before the experiments and the raw data was reported as is, so that comprehensive models can be developed. In our future work we propose a microstructurally motivated model to describe the inelastic behavior of skin and softening that happens to skin and we will investigate mullins effect on skin under biaxial testing. 

\begin{acknowledgements}
The authors gratefully thank Dr. Terry Creasy from Texas A$\&$M University for allowing us to use his facilities.
\end{acknowledgements}


\bibliographystyle{spmpsci} 
\bibliography{reference}


%

\section{Appendix}
In order to compare the Instron cross-head strain and the strain measured by DIC, the strain from grip displacement and the strain from image correlation along the sample from different positions of the sample were calculated, it was seen that the difference between both measurement was less than 5 percent. 
    \begin{equation}
    \varepsilon_d\ = \frac{y_2-y_1}{u_2-u_1}
    \end{equation}

    \begin{figure} [ht]
    \centering
    \includegraphics[scale=0.5]{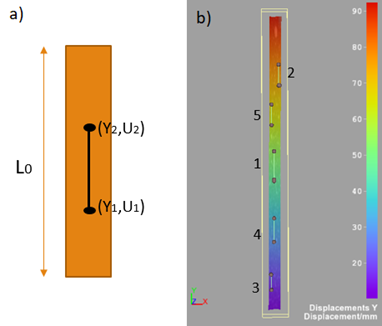}
    \caption{ a) Schematic calculation of Engineering Strain in Y direction from DIC. b) Displacement in Y direction measuring by DIC}
    \label{fig:comp}
    \end{figure}
    
    \begin{table} [ht]
\caption{Both DIC and Instron cross-head engineering strain}
\small
\label{tab:4}
\centering
\begin{tabular}{c|c|c|c}
\hline
\multirow{7}{0.1em}{} Line & $\varepsilon_d$ &  $\varepsilon_c$ & Difference \\
\hline
1 & 1.84  & 1.85 & 0.5 \\
\hline
2 & 1.78 & 1.85 & 3.99 \\
\hline
3 & 1.77 & 1.85 & 4.09  \\
\hline
4 & 1.8 & 1.85 & 2.7 \\
\hline
5 & 1.8 & 1.85 & 2.7  \\
\hline
Average & 1.8 & 1.85 &  2.7 \\
\end{tabular}
\end{table}
Samples with width of 10.4 mm, thickness of 1.65 mm and initial length of 54 mm were tested under simple tension test. The measured engineering strain from crosshead displacement was 1.85 ($\varepsilon_c$). Also, engineering strain was measured by DIC, 5 different lines were chosen from different position of the sample \autoref{fig:comp}. The engineering strain ($\varepsilon_d$) was calculated as the differences between displacements of the top and bottom point of the line ($y_1$ and $y_2$) and divided by the difference between positions of these points on the line ($u_1$ and $u_2$) . Table 1 shows strains measured by both DIC and Instron cross-head. The differences between $\varepsilon_d$ and $\varepsilon_c$ is less than 5 percent on each line and by getting the average of strain on all 5 lines, the difference is 2.7 percent which conforms to the strain measurement by the Instron cross-head.
    
\end{document}